# Optimization of gadolinium oxide growth deposited on Si by high pressure sputtering

Pedro Carlos Feijoo, María Ángela Pampillón, and Enrique San Andrés

Dpto. de Física Aplicada III: Electricidad y Electronica, Facultad de CC. Físicas, Universidad CompIutense de Madrid, Av. Complutense S/N, Madrid 28040, Spain

**Abstract**

High k gadolinium oxide thin layers were deposited on silicon by high-pressure sputtering (HPS). In order to optimize the properties for microelectronics applications, different deposition conditions were used. Ti (scavenger) and Pt (nonreactive) were e-beam evaporated to fabricate metal–insulator–semiconductor (MIS) devices. According to x-ray diffraction, x-ray photoelectron spectroscopy, and Fourier-transform infrared spectroscopy, polycrystalline stoichiometric $Gd_2O_3$ films were obtained by HPS. The growth rate decreases when increasing the deposition pressure. For relatively thick films (40nm), a $SiO_x$ interface as well as the formation of a silicate layer ($GdSiO_x$) is observed. For thinner films, in Ti gated devices the $SiO_x$ interface disappears but the silicate layer extends over the whole thickness of the gadolinium oxide film. These MIS devices present lower equivalent oxide thicknesses than Pt gated devices due to interface scavenging. The density of interfacial defects $D_{it}$ is found to decrease with deposition pressure, showing a reduced plasma damage of the substrate surface for higher pressures. MIS with the dielectric deposited at higher pressures also present lower flatband voltage shifts $DV_{FB}$ in the $C_{HF}$–$V_G$ hysteresis curves. VC 2013 American Vacuum Society. [http://dx.doi.org/10.1116/1.4766184]

I. **INTRODUCTION**

Gadolinium oxide has been widely proposed as an interesting high permittivity dielectric for such applications as blocking oxide in charge-trap flash memory devices and as a third generation of gate dielectrics in CMOS transistors to replace Hf-based oxides and lead the scaling to its final limit.[1–3] This material has drawn attention as high k dielectric based on thermodynamic considerations and good insulating properties: gadolinium oxide has a permittivity value[3] j around 17, a band gap of around 5.3eV, and very good thermal and chemical stability with Si.[2,4] These properties are suitable for the microelectronics requirements.

$Gd_2O_3$ thin films find also other application, such as passivation layers in III–V substrates,[5,6] waveguides,[7] buffer layers in superconductors[8] and scintillating films,[9] protective coatings,[10] etc. Therefore, the study of the growth of this material can also be helpful for these applications.

High k $Gd_2O_3$ films grown on Si substrates by electron beam evaporation,[11] chemical vapor deposition[12] (CVD), low pressure metallorganic chemical vapor deposition[13] (LP– MOCVD), reactive radio frequency sputtering,[14] molecular beam epitaxy[15] (MBE), or atomic layer deposition[16] (ALD) have already been reported, with promising results.[1,17]

This work deals with electrical and structural analysis of gadolinium oxide films grown on Si by means of high pressure sputtering (HPS).[18] This system was demonstrated to be adequate for other high k dielectrics growth.[19–21] During the deposition process, the pressure is maintained in the 0.2–2mbar range, about 3 orders of magnitude higher than in a conventional sputtering system. Thus, the mean free path of the sputtered species in the plasma is very short. This means that the atoms reach the substrate with very low energy (in the order of $k_BT$, $k_B$ being the Boltzmann constant and T the temperature) purely by means of a diffusion process. The low energy of the arriving species prevents the damage of the substrate and the growing film itself.

The purpose of this work is optimizing the HPS growing conditions of $Gd_2O_3$ in order to get good dielectric properties and an interface with low density of defects. Transmission electron microscopy (TEM), glancing incidence x-ray diffraction (GIXRD), x-ray photoelectron spectroscopy (XPS), and Fourier transform infrared spectroscopy (FTIR) were used for studying morphology, composition, stoichiometry, and crystal structure. Metal–insulator–semiconductor (MIS) devices were fabricated with Ti and Pt gates to study the electrical properties and the influence of the choice of the metal electrode.

## II. EXPERIMENT

For structural characterization, gadolinium oxide thin films were grown on 2-in. double side polished n-Si (100) wafers with a resistivity of 200–1000 Xcm. MIS devices were fabricated on single side polished 2-in. n-Si (100) wafers, with a resistivity of 1.5–5.0 Xcm. Before high k $GdO_x$ deposition, substrates were cleaned using a standard Radio Corporation of America (RCA) clean[22] and submerged in a diluted HF solution (1:50) to remove all $SiO_2$ from the surface just before introduction to the sputtering chamber.

$GdO_x$ films were deposited by HPS from a 4.5cm diameter high purity $Gd_2O_3$ target in a pure Ar atmosphere. The 13.54MHz radio frequency (rf) power was 40W. Deposition pressures ranged from 0.50 to 1.3mbar to study the effect of the Ar pressure on the properties of the high k. Base pressure was $1-2 10^6$ mbar, 5 orders of magnitude below working pressure, to minimize contamination. Substrate temperature was maintained at 200C during growth. The deposition time was fixed at 30min to obtain nanometric thick films. Also a thicker sample was grown during 2h at 0.50 mbar to increase signal in the structural characterization. The emission lines of the HPS glow discharge were registered by a Jobin Yvon H-25 monochromator attached to a photon counting system, measuring the plasma emission at wavelengths between 280 and 520nm, with a resolution of 0.1nm. Optical spectra of the plasma at the different Ar pressures were acquired to analyze the changes in the emission of the excited species in the plasma.

For high frequency (HF) capacitance as a function of gate voltage measurements ($C_{HF}$–$V_G$), MIS devices were fabricated. The gate electrodes were 100nm of Ti on one half of each wafer or 16nm of Pt on the other half. Both metals were capped by 200nm of Al, in the Ti case to avoid surface oxidation, and in the Pt case to minimize series resistance and to prevent dielectric perforation while probing. All these metals were deposited by e-beam evaporation with the other half of the wafer shadowed. This way it is assured that for each deposition condition, the high k dielectric is identical and differences can be attributed only to the metal electrode properties. Squares of different sizes ranging from 50·50 to 630·630

um$^2$ were defined by a lift-off procedure, using the photoresist n-LOF 2070. Then, a 300nm thick Al layer was evaporated on the backside to obtain the substrate Ohmic contact. After fabrication and electrical characterization, samples were annealed at 450 C in forming gas (forming gas anneal or FGA process) during 20 min and characterized again electrically and by TEM.

In order to analyze the bonding structure of the films and the interfaces, FTIR spectra of the samples were measured in the 4000–400cm$^1$ region using a Nicolet Magna-IR 750 series II spectrometer working in transmission mode at normal incidence. These spectra were corrected by subtracting the spectrum of a bare Si substrate of the same lot to remove the substrate absorbance. To avoid the native oxide presence on the reference spectrum, the Si substrate was immersed in HF 1:50 for 30s immediately before the FTIR measurement. Thicknesses of the films were extracted from ellipsometry, with a Nanofilm EP$^3$ ellipsometer. MIS devices were physically characterized by means of cross-sectional TEM using a Tecnai T20 microscope from FEI at 200keV. TEM samples were prepared by Dual-Beam. Energy dispersive x-ray (EDX) spectrometry was used during TEM image acquisition to determine the species present in the thin films. GIXRD was measured by a XPERT MRD of Panalytical at x¼0.5 for crystalline structure identification. XPS spectra were obtained with a VG Escalab 200R spectrometer equipped with a MgK$_a$ x-ray source (h ¼1254.6eV), powered at 120W. The background pressure in the analysis chamber was maintained below 210$^8$ mbar during data acquisition. The XPS data signals were taken in increments of 0.1eV with dwell times of 50ms. Binding energies were calibrated relative to the C 1s peak at 284.9 eV. The capacitance and conductance of the MIS capacitors were measured at 100 KHz as a function of the gate bias voltage by an Agilent 4294A impedance analyzer. Density of interfacial defects D$_{it}$ was estimated by the conductance method.[23] The leakage current at accumulation was measured by a Keithley 2636A System, controlled by LABVIEW software. This equipment is able to work simultaneously as a voltage source and an ammeter.

## III. RESULTS AND DISCUSSION

*A. Structural characterization*

In the following, the physical properties of the gadolinium oxide films are studied as a function of the deposition pressure. The rf power was chosen as 40W since it was the rf power used in previous works for the scandium oxide.[18] This power was selected in order to obtain a reasonably high growth rate while minimizing substrate damage.

In Fig. 1, the optical spectra of the Ar plasma sputtering the $Gd_2O_3$ target can be observed in the 385–425nm region for different Ar pressures. This technique was used to determine the optimal growth pressure range of the $GdO_x$ films. No water or $N_2$ is found in the plasma, which is an indication that there are no gas leaks. No oxygen signal is found in the plasma spectra but this fact was also observed by Toledano et al. when sputtering $HfO_2$ in an HPS system.[24] They found that the growing films were fully stoichiometric; thus, this absence should not be an issue. From Fig. 1, it is also clear that the Gd I peaks are under the detection limit for pressures below 0.50mbar (for instance, at 387 and 423nm). Then, for device characterization, the growth pressures were chosen from 0.50 to 1.3mbar in order to guarantee the presence of Gd atoms in the plasma. For higher pressures, the plasma in the HPS chamber is unstable and thus the reproducibility of the growth process cannot be assured.

Once that the pressure range is fixed, the structural characterization of a thick $GdO_x$ film grown on Si at 0.50mbar and 40W during 2h will be discussed. According to ellipsometry, this film has a thickness of 190nm and a refractive index of 1.9. The obtained refractive index is consistent with literature,[25] in the range of 1.8–2.0. The growth rate is about 1.6nm/min.

Figure 2(a) shows the FTIR spectrum in the 1300– 400$cm^1$ region, and it can be compared with the spectrum of pure $Gd_2O_3$ powder shown in Fig. 2(b). In 667$cm^1$, the C–O bond peak appears, due to residual $CO_2$ gas present in the measurement chamber. The valleys at 730 and 610$cm^1$ are related to the crystalline Si vibration modes and are due to the small thickness differences between the sample wafer and the substrate used for correction. The small band at around 1050cm$^1$ is attributed to the transversal optic vibration of the Si-O bond,[26] which should be at 1070cm$^1$ for fully stoichiometric relaxed $SiO_2$. The shift toward lower wavenumbers could be associated with the presence of a Si rich $SiO_x$ or a stressed $SiO_2$ layer.[27,28] The bond formation with Gd, which has a much higher atomic mass than Si, would also shift the Si–O Si band toward lower wavenumbers.[29] Figure 2(a) also shows peaks at around 529 and 451cm$^1$ that can be assigned to the Gd–O vibrations. This can be inferred from direct comparison with the $Gd_2O_3$ powder spectrum shown in Fig. 2(b). Other works also show peaks at 538 and 456cm$^1$ that are associated to Gd–O vibrations in the cubic $Gd_2O_3$ phase.[30–32]

The GIXRD spectrum of the thick unannealed $Gd_2O_3$ layer is depicted in Fig. 3. It clearly shows the most intense diffraction peaks of the cubic $Gd_2O_3$ structure.[33] Since all cubic diffraction peaks are found, with the relative intensities matching the theoretical calculations, these results point to a polycrystalline structure with no preferential growth direction. It is important to note that the GIXRD spectrum shows that the gadolinium oxide is polycrystalline as deposited by the HPS, in other words, without any further annealing.

Summarizing, from these structural measurements it can be concluded that, as is usual with high k material deposition on Si, the 190nm film is composed of a polycrystalline phase of cubic $Gd_2O_3$ that grows with no preferential direction, with a thin $SiO_x$ interfacial layer that is stressed or Si rich.

Once the thick sample has been studied, XPS is used to measure the stoichiometry of the films grown at the different pressures. Figure 4 shows the most relevant peaks in the spectra of the films. The spectra are obtained from the top 2-3nm of the film, which could be an issue if there were surface contamination or water adsorption. To calculate the chemical composition, the peaks at 1187.5 and 531.0 eV were used. They can be attributed to Gd $d_{5/2}$ and O 1s of the $Gd_2O_3$, respectively,[34] with sensitivity factors[35] of 3.41 and 1.61. The sensitivity factor of the Gd $d_{5/2}$ was directly measured from a $Gd_2O_3$ powder sample. The result is that the O/ Gd ratio is 1.560 for all deposition pressures. Then it is concluded that, irrespective of deposition pressure, the films are formed by stoichiometric $Gd_2O_3$.

Finally, TEM images of the MIS capacitors with a Ti gate after the FGA at 450C are shown in Fig. 5. To avoid ambiguities, these TEM samples were prepared from the same devices that were electrically characterized in order to be sure about thicknesses, and thus to obtain an accurate k value. In Fig. 5, it can be observed the $Gd_2O_3$ films that were grown at pressures of (a) 0.50mbar, (b) 0.75mbar, (c) 1.0mbar, and (d) 1.3mbar. First of all, it must be noted that the gadolinium oxide thickness decreases with the deposition pressure. Since all films were deposited for 30min, they show a very low deposition rate (below 1nm/min). Then, an accurate control of the film thickness can be achieved. The decrease in the growth rate is prompted by a reduction of the rate of the rare earth and oxygen atoms that reach the substrate at higher pressures[36] and it has already been observed for $Sc_2O_3$ and $HfO_2$ in previous works.[18,19]

In Fig. 5(a), the high k dielectric grown at 0.50mbar presents a polycrystalline $Gd_2O_3$ layer that is on top of an amorphous gadolinium silicate ($GdSiO_x$). A 3.8nm $SiO_x$ interfacial layer appears between the $GdSiO_x$ and the Si substrate. The composition of these thin layers was measured by EDX spectrometry during TEM images acquisition. The presence of an amorphous gadolinium silicate means that gadolinium oxide is reacting with the underlying $SiO_2$/Si. It cannot be assured if this silicate was formed during deposition or during annealing since there are no TEM images before the FGA. However, XRD did not show any $GdSiO_x$ peaks in the unannealed samples, so this reaction probably happened during the FGA. The instability of $Gd_2O_3$ against silicate formation on Si substrates has already been reported.[37,38] This low temperature silicate formation could be interesting for device applications, in order to keep $SiO_2$like properties while having $Gd_2O_3$-like permittivities, as was thoroughly studied in the HfSiON system.[39–41] The presence of a top poly-$Gd_2O_3$ agrees with XRD results, suggesting that before FGA the sample was polycrystalline.

For pressures above 0.5mbar, the interfacial $SiO_x$ seems to be completely scavenged by the Ti gate[42] or dissolved inside the $GdSiO_x$.[43,44] Figures 5(b) and 5(c) indicate that for deposition pressures of 0.75 and 1.0mbar, the dielectric stack also consists of an amorphous $GdSiO_x$ layer and a poly-$Gd_2O_3$ layer whose thicknesses decrease with deposition pressure. Figure 5(d) shows that for 1.3mbar, only an amorphous $GdSiO_x$ film is found. This means that the sample was so thin initially that the whole film has become gadolinium silicate. Since no remaining $Gd_2O_3$ film is present, no information on crystallinity can be obtained. However, it is reasonable that, as for the other pressures, the grown film was initially polycrystalline. Finally, it is relevant to note that no Ti signal was found by EDX inside the films, indicating that there is no Ti diffusion during FGA.

*B. Electrical characterization*

Figure 6 shows $C_{HF}$–$V_G$ curves of the thinner MIS devices where the dielectric was grown at 1.3 mbar during 30 min. The curves are represented before and after the FGA and for Pt and Ti gate electrodes. First, it must be noted that before the FGA, the capacitance curve is almost identical for both electrodes. This happens in all samples; only on the 0.75mbar sample, there are differences between Ti and Pt, which points to a higher thermal budget in this sample during Ti evaporation due to excessive deposition rate. Excluding this sample, these results point to the interpretation that the metal evaporation does not influence the underlying $Gd_2O_3$. Previous works[45] on plasma oxidized $Gd_2O_3$ found that the heating of the sample during Ti evaporation always induce some interface scavenging. However, for $Gd_2O_3$ samples in this work, no evaporation-induced scavenging is found. In both cases, the thickness of the film is similar, so the difference can be the crystallinity of the films: the plasma oxidized $Gd_2O_3$ was amorphous while this case presents a polycrystalline layer. Also the flatband voltage of the curves are similar for both electrodes (0.33V for Ti and 0.38V for Pt), and there should be a 1.4V difference due to the workfunction difference [4.3eV for Ti and 5.7eV for Pt (Ref. 46)]. Since the underlying $Gd_2O_3$ is identical in both cases, this effect can be due to electrode contamination during evaporation (coming from the crucible, molybdenum for Ti or graphite for Pt) or most likely to dipole formation in the $Gd_2O_3$/metal interface.[47]

In Fig. 6(a), it can be observed that the annealing process does not modify the curve when the top metal is Pt, with only a very small variation in flatband voltage (0.38V for the unannealed, 0.34V after the FGA). This is the result of Pt being a noble metal, which should not react with the dielectric. Thus, no changes in the devices are inferred from the $C_{HF}$–$V_G$ curve. On the other hand, as it can be seen in Fig. 6(b), for the Ti gated device the capacitance in accumulation more than triples after the FGA, going from 0.55 up to 1.8lF $cm^2$. This means that the Ti gate is scavenging the $SiO_x$ interface during the FGA, increasing the effective permittivity of the dielectric stack and thus decreasing the equivalent oxide thickness (EOT). In fact, the EOT decreases dramatically from 6.2 to 1.4nm in this sample. There is also a flat-band displacement toward lower voltages, which can be due to fixed charges in the dielectric or a dipole change in the Ti/$GdSiO_x$ interface.

The EOTs for all samples are represented in Fig. 7 before (a) and after the FGA at 450 C (b). It can be observed that before FGA the EOT of the devices is high, above 5nm for 1.3mbar, and even higher for the lower deposition pressures. This is consistent with the thicknesses of the grown films obtained by TEM, together with the interfacial $SiO_x$ interfacial layer. On the other hand, for all Pt gated devices, the EOT does not change significantly with the FGA, while Ti gated devices show a noticeable EOT reduction. As it was seen above, this can be related to the interface scavenging properties of the Ti metal gate. Only the device fabricated at 0.50mbar does not show this EOT reduction. Since interface scavenging requires oxygen diffusion through the high k film,[48] for the thicker layer no scavenging takes place (in fact, there is some interfacial regrowth as the increase in EOT values after annealing shows).

The effective permittivity $k_{eff}$ of the grown films can be calculated from the following equation:

$$k_{eff} = \frac{k_{SiO}}{EOT} t_{dielectric} \quad (1)$$

where $k_{SiO}$ is the dielectric permittivity of the silicon dioxide (3.9) and $t_{dielectric}$ is the total thickness of the dielectric stack obtained by the TEM images. In this calculation, the gadolinium silicate, gadolinium oxide, and interfacial $SiO_2$ are averaged, thus the permittivity is an effective value. However, this is the relevant factor from a device point of view, since even a very high k value would be compromised for a too thick $SiO_x$ interfacial layer. The results of the Ti gated MIS devices are presented in Table I. In spite of the dispersion, a k value above 11 is obtained for the samples where the $SiO_x$ interface has been completely scavenged. This value is lower than $Gd_2O_3$ bulk relative permittivity (17, Ref. 3), but this fact is usual for many thin film materials. In this case, the silicate formation observed in TEM is the most likely reason of the decrease in the effective permittivity. In any case, the value obtained for the $GdSiO_x$, the sample deposited at 1.3 mbar, is very close to the bulk $Gd_2O_3$ value, which suggests that the silicate is Gd rich (in other words, the initial $SiO_2$ was very thin).

The density of interfacial defects $D_{it}$ before and after the FGA is estimated by the conductance method and is represented as a function of the deposition pressure in Fig. 8. Before FGA higher deposition pressures give rise to better quality interfaces, with $D_{it}$ in the order of $1–2 10^{10}$ $eV^1$ $cm^2$ for the higher pressures. Then, the substrate surface damage is lower during growth for higher deposition pressures. This is due to the shorter thermalization lengths of the species in the plasma that reach the substrates purely by means of a diffusion process. Figure 8 also shows an unexpected increase in the density of interfacial defects after the FGA, for both Ti and Pt gated devices. For the Ti gated devices, this increase should not be surprising since the scavenging reduces the $SiO_x$ in the interface, increasing the number of defects. However, it also happens for the Pt gated devices, which suggests that the origin of the increase is in the silicate formation observed by TEM. The silicate grows by dissolving the $SiO_x$ layer or even reacting with the Si substrate, which would leave an interface with a higher density of defects. For device applications, these high defect densities should be addressed.

In addition to this, it is found that only the Gd$_2$O$_3$ grown at 0.50 mbar presents a significant hysteresis while films grown at higher pressures do not show any flatband voltage shifts ΔV$_{FB}$, as can be seen in the C$_{HF}$–V$_G$ hysteresis curves shown in Fig. 9. The curves were acquired starting in accumulation. This means that for the 0.50mbar sample, there is a positive charge trapping when the device is in inversion. Then higher deposition pressures produce thermalization of the plasma species during deposition and thus the dielectric film grows in a less aggressive environment. Therefore, higher deposition pressures mean a lower number of bulk defects and also a lower density of interfacial defects.

Finally, the leakage current was measured in accumulation. The leakage current of the Pt gated MIS devices were under the detection limit ($10^{13}$ A), even for the largest areas and the thinnest dielectrics. These low leakages are probably caused by the SiO$_x$ thick film at the dielectric/Si interface. The J$_G$–V$_G$ curves of the Ti gated devices are represented in Fig. 10. For deposition pressures below 1.0mbar, the dielectrics are so thick that they present low current density (below $10^8$ A cm$^2$). Only the 5nm GdSiO, which was deposited at 1.3mbar, shows a significant leakage current, reaching values of $10^4$ A cm$^2$ at 1V and breaking down at 1.6V. These low levels of leakage are due to the high thickness of the films.

IV. **SUMMARY AND CONCLUSIONS**

This work demonstrates the viability of producing thin Gd$_2$O$_3$ and GdSiO$_x$ films by HPS on Si as high k dielectrics for CMOS applications. It analyzes structural and electrical characterization of Gd based oxide films for different conditions of deposition pressure and the influence of the choice of the top electrode metal, comparing a nonreactive metal (Pt) and an interface scavenger (Ti).

A thick sample was used to confirm the growth of the gadolinium oxide and a polycrystalline Gd$_2$O$_3$ phase was found. In thinner high k films, it was observed that Gd$_2$O$_3$ react with the Si substrate and an amorphous silicate is formed, probably during the FGA at 450C. While the noble metal Pt does not react with the dielectric in the MIS devices, the Ti gate scavenges the SiO$_x$ interfacial layer for thin Gd$_2$O$_3$ films, decreasing dramatically the EOT of the stack, with low leakage current densities. Higher deposition pressures showed a higher quality in the high k/Si interfaces with lower densities of defects and hysteresis, although the silicate formation during FGA also degrades the interface quality. An effective permittivity above 11 was calculated for the high k dielectric stacks.

**ACKNOWLEDGMENTS**

The authors would like to acknowledge "C.A.I. de Tecnicas Físicas," "C.A.I. de Rayos X," and "C. A. I. de Espectroscopía y Espectrometría" of the Universidad Complutense de Madrid. The "Instituto de Nanociencia de Aragon" is also gratefully acknowledged for the sample preparation and the TEM image acquisition. The authors also thank the "Instituto de Catálisis y Petroleoquímica" for the XPS measurements. This work was funded by the Spanish Ministerio de Economía y Competividad through the project TEC2010-18051. M. A. Pampillon and P. C. Feijoo works were funded by FPI Grant BES-2011-043798 and FPU Grant AP2007-01157, respectively.

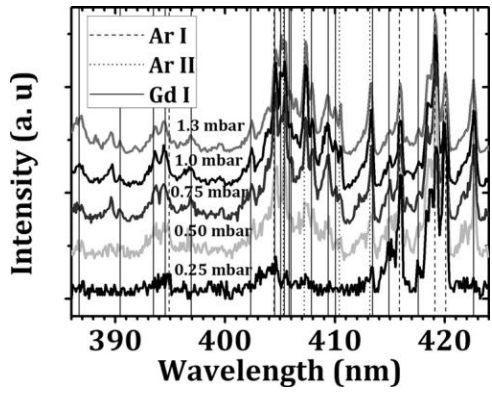

FIG. 1. Plasma spectra of $Gd_2O_3$ sputtered at different Ar pressures and 40W of rf power. The most important peaks of Gd I, Ar I, and Ar II are denoted by vertical lines. Gd I peaks are not detected for pressures below 0.50mbar.

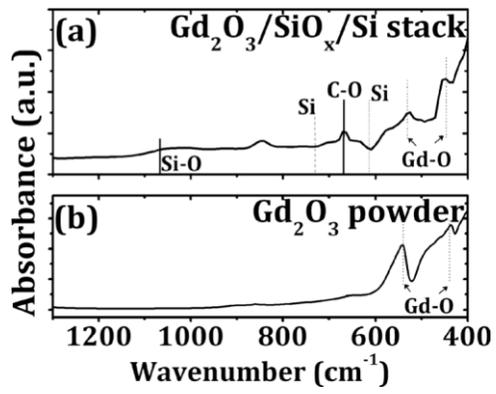

FIG. 2. (a) FTIR spectrum of the 190nm thick $Gd_2O_3$ on Si after substrate correction. (b) FTIR spectrum of $Gd_2O_3$ powder.

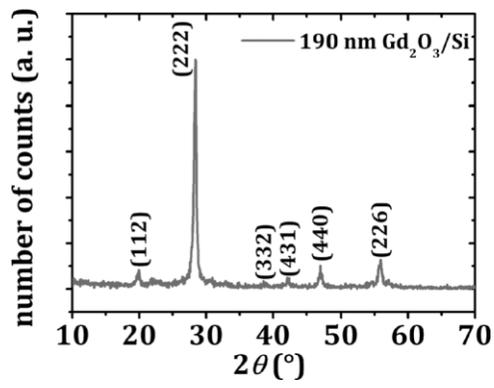

FIG. 3. GIXRD spectrum of 190nm $Gd_2O_3$ on Si. The most relevant peaks of the cubic $Gd_2O_3$ structure are shown.

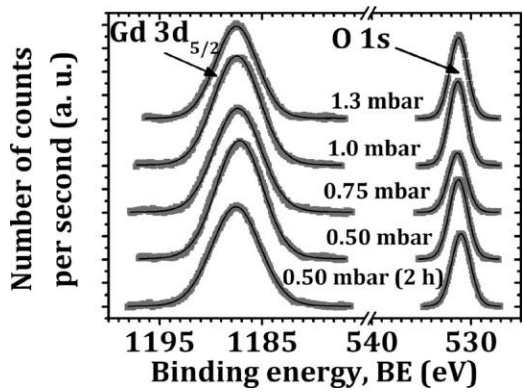

FIG. 4. XPS spectrum of $Gd_2O_3$ on Si. The atomic factor of the Gd $3d_{5/2}$ peak is 3.41 and the O 1s peak is 1.61. The atomic ratio [O]/[Gd] is 1.5; then, the oxide is stoichiometric ($Gd_2O_3$).

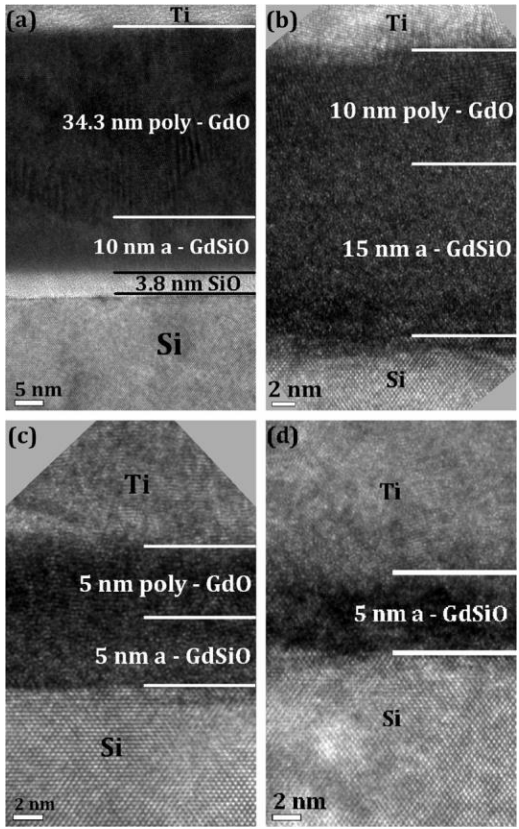

FIG. 5. TEM images of Ti/GdO/Si MIS devices after FGA at 450C. Gadolinium oxide was deposited on Si by HPS at 0.50mbar (a), 0.75mbar (b), 1.0mbar (c), and 1.3mbar (d) at 40W during 30min. No $SiO_x$ interlayer is observed for pressures above 0.50mbar due to Ti scavenging. An amorphous gadolinium silicate is formed. For pressures below 1.0mbar, a polycrystalline phase is observed.

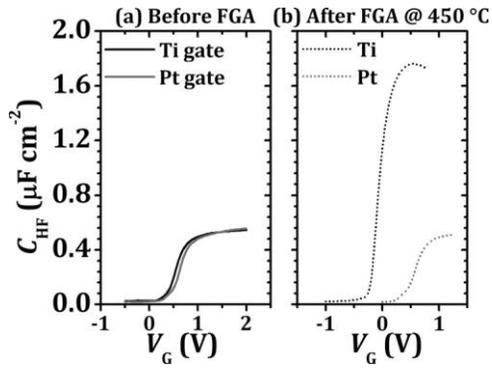

FIG. 6. $C_{HF}$–$V_G$ characteristics at 100kHz for samples before (a) and after (b) the FGA at 450C with Ti and Pt gates. For the Pt electrode, the curve does not change significantly after FGA. For the Ti electrode, the capacitance strongly increases due to interface scavenging.

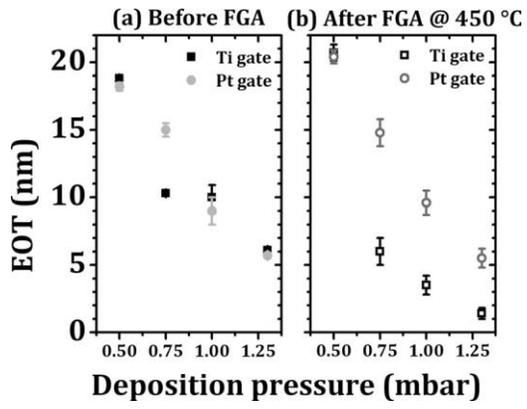

FIG. 7. EOT as a function of deposition pressure before (a) and after (b) the FGA at 450C.

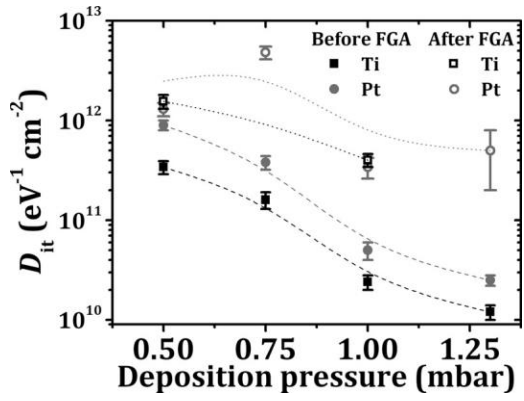

FIG. 8. Density of interfacial defects $D_{it}$ in function of the deposition pressure estimated by the conductance method. The quality of the interface is increased with the deposition pressure. Lines were drawn as a guide to the eye.

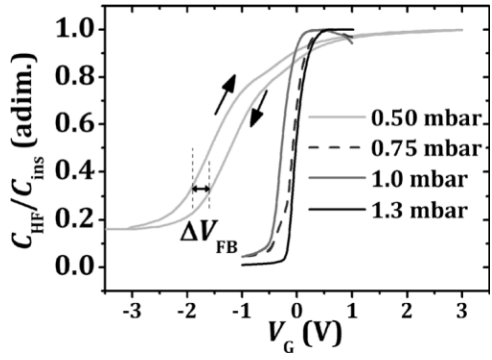

FIG. 9. $C_{HF}$–$V_G$ hysteresis curves for Ti/GdO/Si stacks, with $Gd_2O_3$ HPS deposited at several pressures. For pressures above 0.75mbar, flatband voltage shift $DV_{FB}$ is negligible.

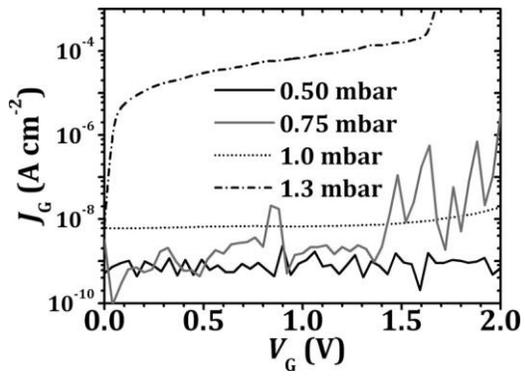

FIG. 10. Leakage density current for Ti/GdO/Si stacks. Leakage is only appreciable for the 5nm thick GdSiO dielectric.

TABLE I. Deposition pressure, thickness, and electrical parameters of the Ti gated MIS devices, and effective permittivity of the dielectric stack.

| Deposition pressure (mbar) | Dielectric thickness (nm) | Dielectric stack | EOT (nm) | Effective relative permittivity (adimensional) |
|---|---|---|---|---|
| 0.50 | 49.1 | $Gd_2O_3/GdSiO_x/SiO_x$ | 20.7 | 9 |
| 0.75 | 25 | $Gd_2O_3/GdSiO_x$ | 6.1 | 16 |
| 1.0 | 10 | $Gd_2O_3/GdSiO_x$ | 3.5 | 11 |
| 1.3 | 5 | $GdSiO_x$ | 1.4 | 14 |